\documentclass[aps,prl,superscriptaddress,floatfix,twocolumn,showpacs]{revtex4-1}

\def\xv{{\bf x}}

\def\Fv{{\bf F}}

\def\Mv{{\bf M}}

\def\be {\begin{equation}}
\def\ee {\end{equation}}
\def\ber {\begin{eqnarray}}
\def\eer {\end{eqnarray}}
\def\nn {\nonumber}
\def\nablabold{\mbox{\boldmath $\nabla$}}

\usepackage{graphicx,graphics}
\usepackage{dcolumn}
\usepackage{amsmath,amssymb,amsfonts}
\usepackage{latexsym,verbatim}
\usepackage{bm}
\usepackage{color}
\usepackage[breaklinks=true,colorlinks,citecolor=blue,linkcolor=blue,urlcolor=blue]{hyperref}

\usepackage{fancybox}
\usepackage{tabularx}


\begin{document}

\author{Erica E. Hroblak }
\affiliation{Department of Physics and Astronomy, University of Missouri, Columbia, Missouri 65211,~USA}

\author{Alessandro Principi}
\affiliation{Radboud University, institute for Molecules and Materials, NL-6525 AJ Nijmegen, The Netherlands}

\author{Hui Zhao}
\affiliation{Department of Physics and Astronomy, University of Kansas, Lawrence, Kansas 66045,~USA}

\author{Giovanni Vignale}
\affiliation{Department of Physics and Astronomy, University of Missouri, Columbia, Missouri 65211,~USA}

\title{Negative electronic compressibility enables electrically-induced charge density waves in a two-dimensional electron liquid} 

\begin{abstract}
We show that the negative electronic compressibility of two-dimensional  electronic systems at sufficiently low density enables the generation of charge density waves through the application of a uniform force field, provided no current is allowed to flow.  The wavelength of the density oscillations is controlled by the magnitude of the (negative) screening length, and their amplitude is proportional to the applied force. Both are electrically tunable.    
\end{abstract}


\maketitle

{\it Introduction} -- The occurrence of negative compressibility is a peculiar feature of electronic systems, whose stability against long-range Coulomb repulsion is ensured by the presence of a background charge, such as ionized atomic cores in metals, ionized dopants, or gates in semiconductors.\cite{GV}  At moderately low density, the chemical potential of the electrons decreases with increasing density,
implying a negative compressibility (see Fig. 1).
This happens when the negative exchange and correlation contributions to the energy, arising from the electron-electron interaction,  dominate over the positive kinetic energy -- which is inevitable at sufficiently low density.  \cite{GV,Bello,Kusminskiy,Skinner,Steffen}     
In an ordinary system, a negative compressibility would be a sign of instability, leading to collapse or phase separation,  but the electron liquid is generally protected against such instabilities by its background charge.  It is only at extremely low densities or at very high magnetic fields that non-uniform phases such as the Wigner crystal~\cite{Wigner} or stripe and bubble phases~\cite{Fogler} are expected to occur.

Experimentally, a negative electronic compressibility has been observed as a positive quantum mechanical correction to the classical capacitance of a capacitor, whose plates are two-dimensional electron layers formed in a semiconductor quantum well (GaAs)\cite{Eisenstein}, a carbon nanotube~\cite{Ilani}, or the interface between two oxides (LaAlO$_3$/SrTiO$_3$)~\cite{Li}.  The effect has also been demonstrated in the two-dimensional electron gas at high magnetic field~\cite{Kravchenko} and in graphene, also at high magnetic field~\cite{Skinner2}.   Very recently,  the decrease of the chemical potential with increasing density has been directly observed by ARPES spectroscopy in two-dimensional monolayers of WSe$_2$~\cite{Riley} and quasi-three dimensional spin-orbit correlated materials~\cite{He},  by conductivity measurements in graphene-MoS$_2$ heterostructures,~\cite{Larentis} and by capacitance measurements  in graphene-terminated black phosphorous heterostructures.~\cite{Wu}  In these experiments the electronic densities are considerably larger than in conventional semiconductor heterostructures, ruling out the spontaneous occurrence of inhomogeneous phases.  
\begin{figure}[h]
\begin{center}
\includegraphics[scale=0.35]{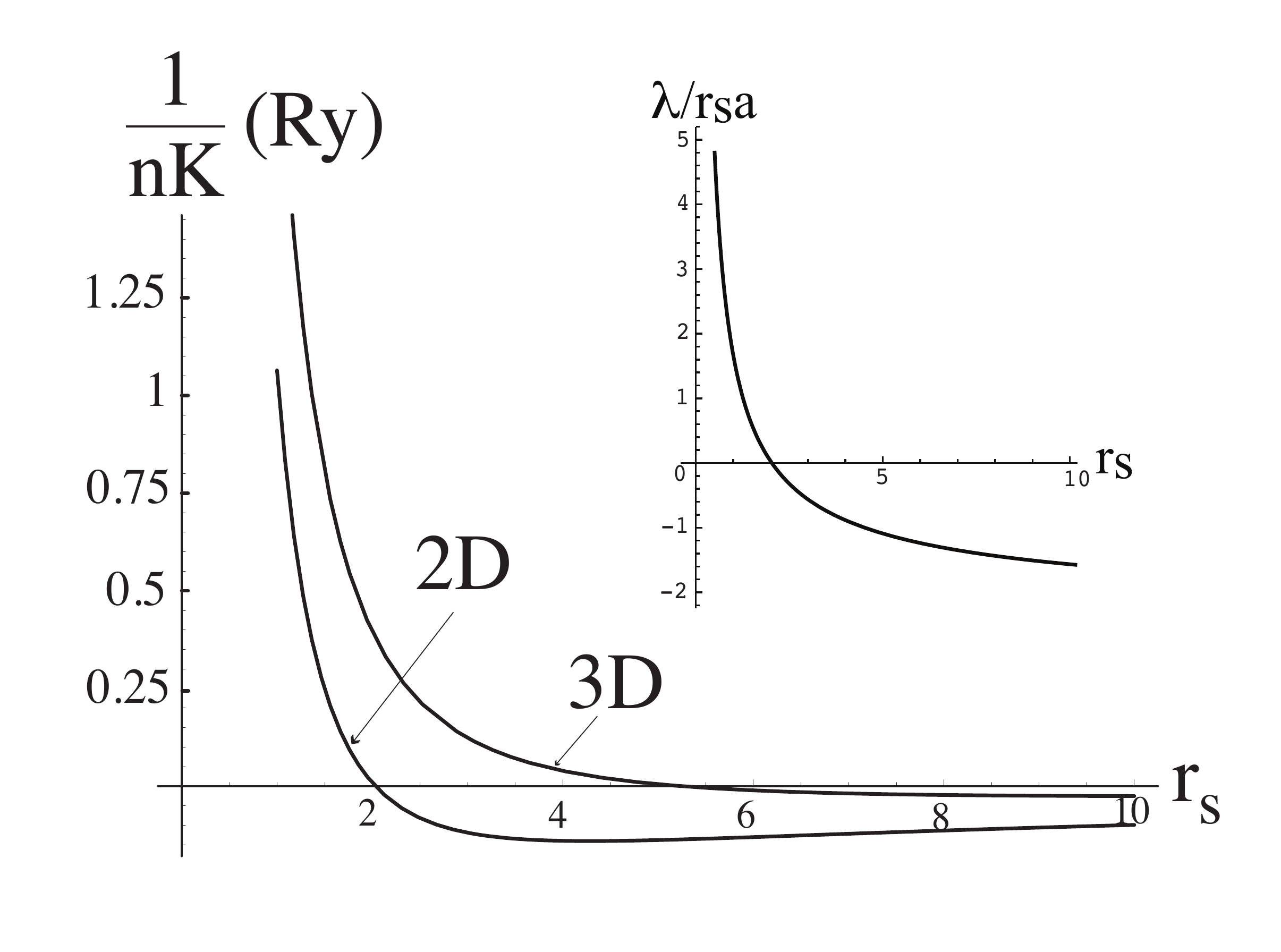}
\end{center}
\caption{Inverse electronic compressibility $K$ over density $n$ as a function of the electron gas parameter $r_s = \left(\frac{3}{4 \pi n }\right)^{1/3}a^{-1}$ in 3 dimensions and $r_s = \left(\frac{1}{\pi n }\right)^{1/2}a^{-1}$ in 2 dimensions, where $a$ is the effective Bohr radius. The Rydberg unit (Ry) is $e^2/(8 \pi \varepsilon a)$. Adapted from Ref.~\onlinecite{GV}.  Inset: the two-dimensional screening length $\lambda = \frac{4\pi\varepsilon}{e^2}\frac{\partial \mu}{\partial n}$ in units of $r_sa$ vs. $r_s$.}  
\end{figure}

In this paper we introduce a new context in which a negative electronic compressibility can be used:  the controlled generation of charge density waves.  Charge density waves (CDWs) are static oscillations of the conduction charge density.\cite{Overhauser,Thorne}  They have long been studied as a possible alternative to incoherent single-particle transport.\cite{Fukuyama78,Lee79}  Unfortunately, the very same charge background that allows the compressibility to go negative is a major obstacle to the formation of charge density waves due to the electrostatic energy cost that such waves would incur.  However, we will now show that a uniform and steady force applied to the electrons when the  compressibility is negative does produce a charge density wave, provided the electrons are not allowed to flow.   One way to apply such a steady force is to pass a current through an adjacent electron layer:  in this setup, schematically shown in Fig. 2,   the force arises from the Coulomb drag effect, whereby momentum is steadily transferred from a current-carrying layer  to a non-current-carrying one.\cite{Pogrebinski,Price,Gramila,Rojo,Narozhny} Another way to achieve the result is to drive a current in a single layer in the presence of a perpendicular magnetic field and let the Hall effect generate the required force in the direction perpendicular to the current. 
CDWs not aligned with the applied current can be generated in Coulomb-drag setups in the presence of a magnetic field, or if the passive layer is a gapped topological insulator, as a consequence of the (standard or anomalous) Hall drag effect.~\cite{Liu_arxiv_2016}
In any case the CDW appears when the two-dimensional screening length $\lambda = \frac{4\pi\varepsilon}{e^2}\frac{\partial \mu}{\partial n}$  ($K^{-1}=n^2\partial \mu/\partial n$) is negative, which occurs for $r_s>2$ (see inset of Fig. 1). The wave vector of the CDW is parallel to the direction of the force,  and its wavelength is $|\lambda|$.   Both the amplitude  and the wavelength of the CDW can be electrically tuned: the former by changing the applied force, the latter by changing, via a gate, the density of the passive layer, and hence the value of $\lambda$.

\begin{figure}[h]
\begin{center}
\includegraphics[scale=0.75]{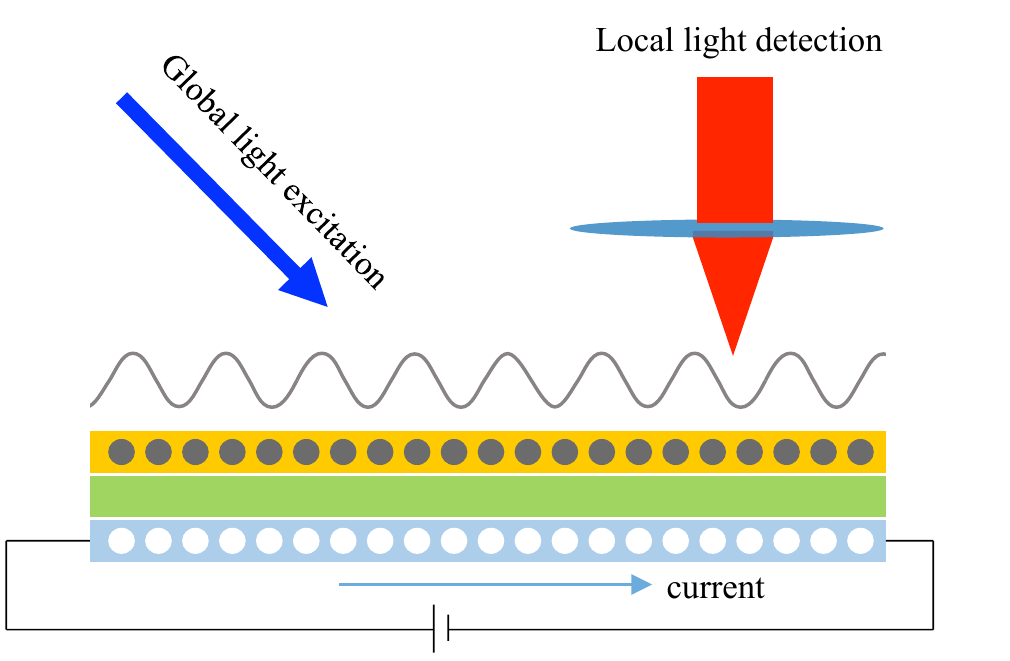}
\end{center}
\caption{Schematic setup for the observation of electrically generated charge density waves.  The current in the lower layer creates, via Coulomb drag, an electric field in the top layer, in which no current flows.  When the compressibility of the electron gas in the top layer is negative, the force exerted by this electric field results in the formation of a charge density wave with a wavelength determined by the absolute value of the compressibility. The density modulation can be detected by optical methods, such as differential absorption or diffraction.}
\end{figure}

{\it Model and its solution} -- We consider a two-dimensional electron gas of uniform two-dimensional density $n$ on an infinite strip of length $L$ in the $x$ direction.  A uniform background of positive charge exactly neutralizes the electron charge.  We now assume that a force $\Fv = F\hat\xv$ acts on each electron.  In the Coulomb drag set-up of Fig. 2 $\Fv$ is proportional, via the transresistance~\cite{Gramila}, to the current that flows in the adjacent layer. We begin by considering the classical equilibrium solution.  Because no current can flow in the $x$ direction, the external force must be exactly balanced by the electric field that arises from the rearrangement of the charge in the plane.  Assuming that the density remains uniform in the $y$ direction the equilibrium condition is
\be\label{ClassicalEquilibrium}
F+\int_{-L/2}^{+L/2}dx'~ \frac{e^2 \delta n(x')}{2\pi\varepsilon (x-x')}=0\,,
\ee
where $\delta n(x')$ is the deviation of the two-dimensional electron density from equilibrium.  It is natural to express $x$ in units of $L/2$  and the density in units of $\ell^{-2}=\frac{2\pi\varepsilon F}{e^2}$.  In these units the solution of Eq.~(\ref{ClassicalEquilibrium}) is easily seen to be
\be
 \delta n(x)=\frac{T_1(x)}{\pi\sqrt{1-x^2}}\,,
\ee
where $T_1(x)=x$ is the Chebyshev polynomial of order $1$. Notice that this solution does not depend on any microscopic length scale: it simply scales with the geometric size.

As the size of the system decreases,  it becomes essential to include the quantum mechanical force arising from the gradient of the chemical potential: $\Fv_q=-\nablabold \mu$.  In the local density approximation this is approximated as
\be
\Fv_q=-\left.\frac{\partial \mu}{\partial n}\right\vert_{n} \nablabold \delta n\,.
\ee
{\bf Strictly speaking, the local density approximation is valid when the scale of variation of the density is large in comparison to the average inter particle spacing.  While this  condition is only marginally satisfied, we will argue in the following that it is nevertheless adequate to predict density oscillations in the 2DEG, provided the wave vector remains smaller than 2$k_F$, where $k_F=\sqrt{2\pi n}$ is the Fermi wave vector.}   By further assuming that we are in the linear response regime, we evaluate the derivative of $\mu$ at the homogeneous equilibrium density.  In terms of the dimensionless screening length, 
\be
\bar \lambda=\frac{\lambda}{L}=\frac{4\pi\varepsilon}{Le^2}\frac{\partial \mu}{\partial n}\,,
\ee
the equilibrium condition becomes
\be\label{IntegroDifferential}
1+\int_{-1}^{+1}dx'~ \frac{\delta n(x')}{x-x'} -\bar \lambda \frac{d \delta n(x)}{dx}=0\,.
\ee

This linear integro-differential equation can be solved numerically by expanding the density in Chebyschev polynomials as follows:
\be
\delta n(x)=\sum_{j=1}^\infty c_j T_{2j-1}(x)\,,
\ee
where only odd polynomials are included, consistent with the symmetry of the problem.   We than have
\be
\frac{d \delta n(x)}{dx}=\sum_{j=1}^\infty c_j (2j-1)U_{2j-2}(x)\,,
\ee
where $U_n(x)$ is the associated Chebyschev polynomial.
Substituting these expressions into Eq.~(\ref{IntegroDifferential}), multiplying both sides by $\sqrt{1-x^2} U_{2k-2}(x)$, with $k \geq 1$, and integrating over $x$ with the help of standard integrals for the Chebyschev polynomials, we arrive at a set of linear algebraic equations for $c_j$:
 \be
\sum_{j=1}^\infty M_{kj} c_j=\delta_{k1}\,,~~~k=1,2...
\ee
where
\ber\label{MIJ}
M_{kj}&=&2\left\{\frac{1}{1-4(k+j-1)^2}+\frac{1}{1-4(k-j)^2}\right\}\nn\\
&+&\bar \lambda (2 k-1)\delta_{kj}\,.
\eer
The solution is
\be
\delta n(x)=\sum_{j=1}^\infty \left[\Mv^{-1}\right]_{j1}T_{2j-1}(x)\,,
\ee
where $\Mv^{-1}$ denotes the inverse of the matrix $\Mv$, whose matrix elements are given by Eq.~(\ref{MIJ}).
The solution is obtained by numerically inverting $\Mv$ on a sufficiently large set of Chebyshev polynomials, such that the results become independent of basis size.

The character of the solution depends dramatically on the sign of  $\lambda$.
When $\lambda$ is positive, which is the normal state of affairs when the equilibrium density  is high, the correction to the classical solution is hardly observable.  The space charge is essentially excluded from the center of the system and accumulates against the edges as expected (Figure 3).  
\begin{figure}[ht]
\begin{center}
\includegraphics[scale=.3]{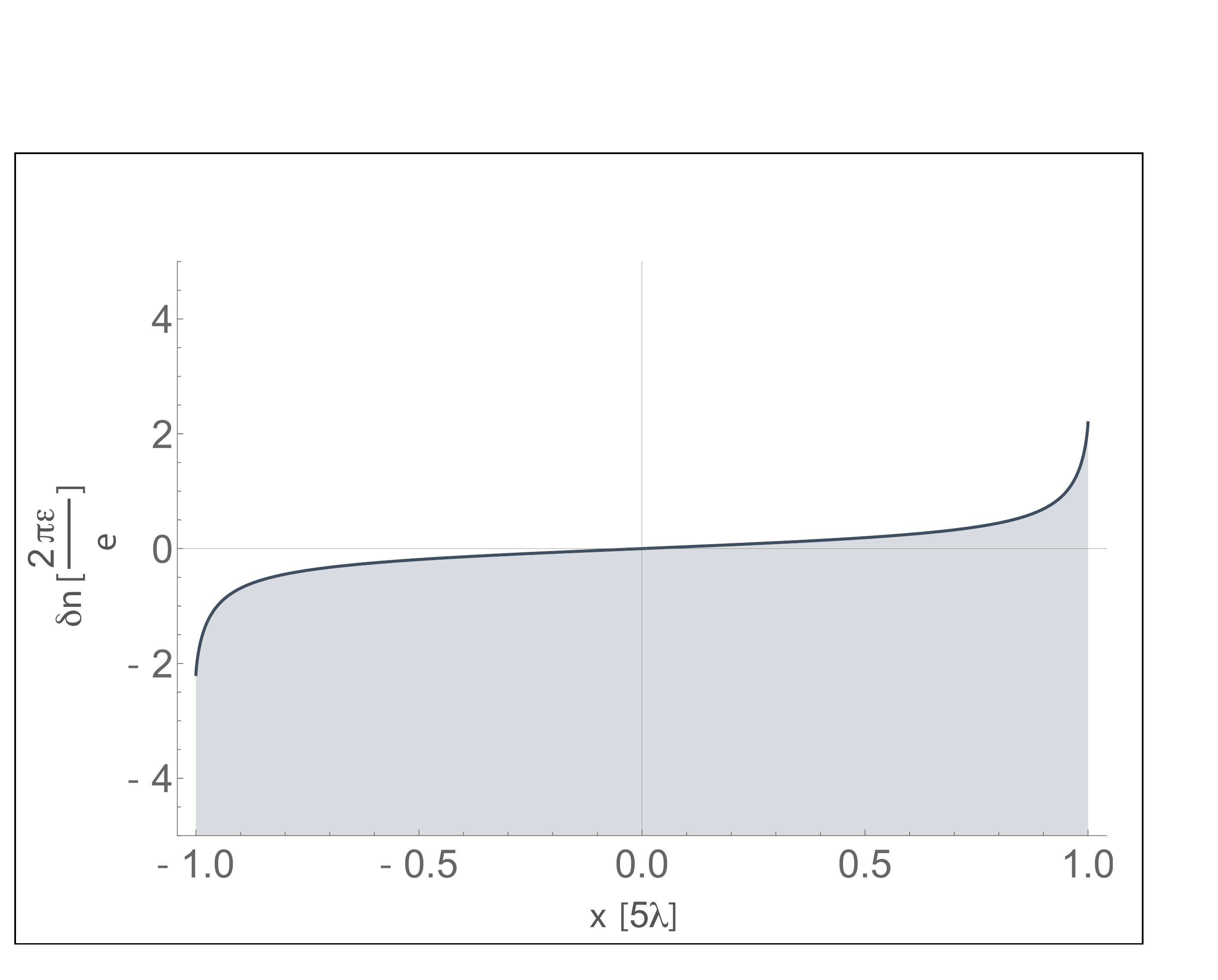}
\end{center}
\caption{The equilibrium solution to Equation (5) for a positive compressibility, $\lambda>0$. The magnitude of the force due to the externally applied field is set to $F_{ext}=1 \frac{eV}{m}$ and the density $n$ is expressed in units of $\frac{2\pi\varepsilon}{e}\approx 3.48 \times10^4~cm^{-2}$.  The length of the bar is taken to be $L=10\lambda$.}
\end{figure}

\begin{figure}[ht]
\begin{center}
\includegraphics[scale=.3]{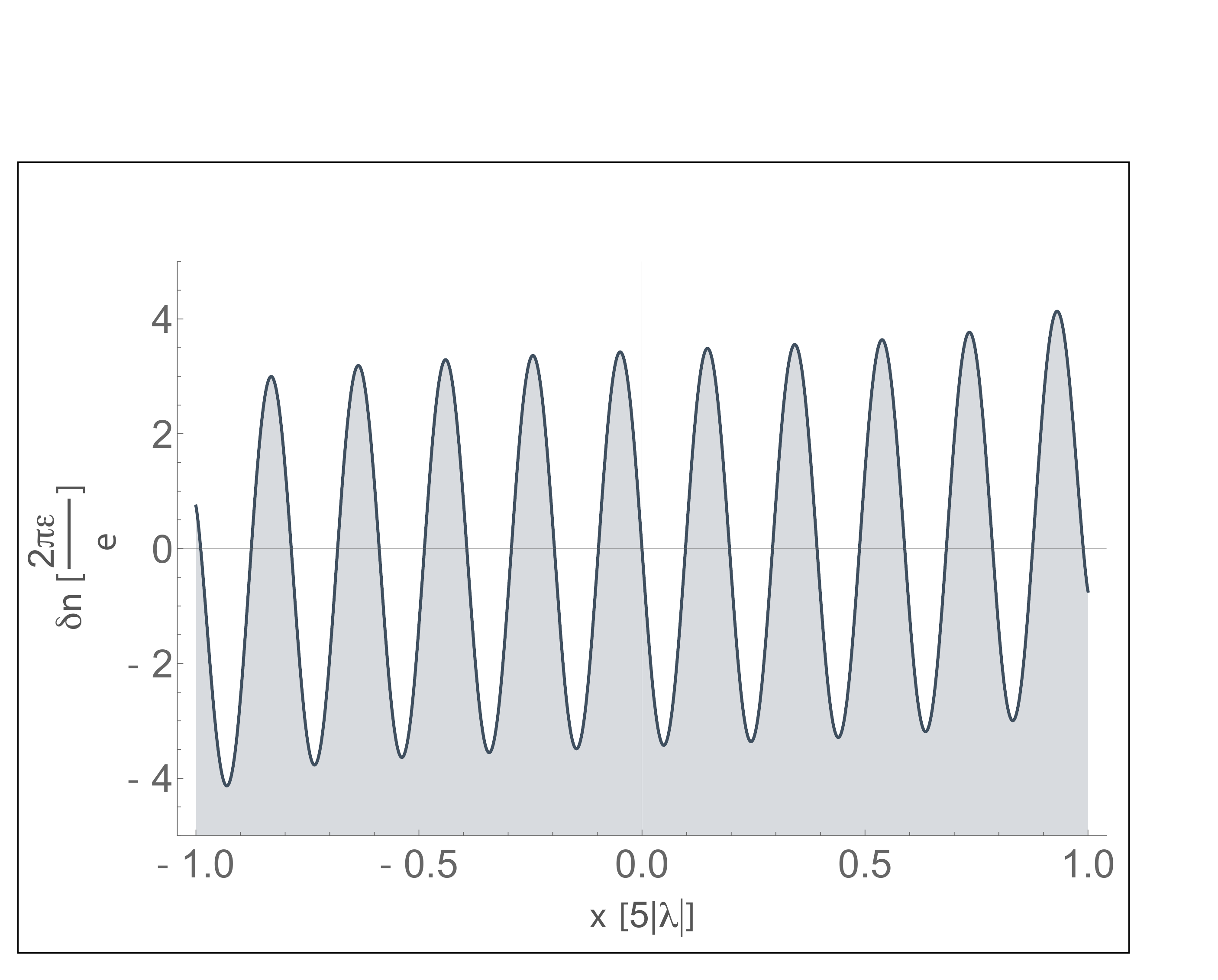}
\end{center}
\caption{The equilibrium solution to Eq.~(\ref{IntegroDifferential}) for a negative compressibility, $\lambda<0$.  The length of the bar is taken to be $L=10|\lambda|$.}
\end{figure}

When the equilibrium density is low, however, the exchange and correlation energies overwhelm the kinetic energy \cite{Steffen,Wigner} leading to a negative value for $\lambda$. Figure (3) shows the surprising result of the simple change of sign of $\lambda$. Instead of accumulating on the edges as in the classical picture, charges distribute along the width of the layer in a sinusoidal pattern. The amplitude and frequency of these oscillations depend on the applied force, the width of the layer, and the value of $\lambda$, providing ample opportunity for electronic tuning. 
{\bf In the central region of the bar ($|x|\ll 1$) an analytic solution of the equation can be obtained and it is given by the sum of the classical equilibrium solution and a simple oscillation of wavelength $|\lambda|$.  The details of the analytic solution are supplied in the Appendix. Remarkably, the amplitude of the density oscillations is a non-analytic function of $|\lambda|$, going as $1/2\cos(\pi/|\bar\lambda|)$ (see Appendix).  This means that, for a given magnitude of the force, the amplitude of the density oscillations does not vanish in the mathematically equivalent limits of $\lambda\to 0$, or $L\to \infty$.  But while these two limits are mathematically equivalent, their physical significance is entirely different.  In the limit $\lambda \to 0$, with finite $L$, the local density approximation breaks down due to rapid density variation and the solution is not expected to have a physical significance.  Whereas, in the limit of large $L$ and finite $\lambda<0$ our solution is expected to be physically meaningful and independent of system size, provided $\lambda>1/(2k_F)$.  

In a two-dimensional electron gas the Fermi wave vector is related to the average inter particle distance by $k_F = \sqrt{2}/(r_s a)$\cite{GV}.
Looking at the inset of Fig. 1 we see that the key quantity $2 k_F|\lambda| =2 \sqrt{2}|\lambda|/(r_sa)$ is never much larger than 1 -- rather, it approaches a limiting value $\sim 5.6$ in the  limit of large $r_s$.
This casts some doubts on the validity of the local density approximation.  A more accurate solution of the problem, including full non locality, remains therefore an important issue to be addressed in future work.  For the time being, we observe that the static density-density response function of a 2DEG is known to be a fairly constant function of wave vector  equal to $d\mu/dn$ for all $q$'s up to $2k_F$ (see ref.~\onlinecite{GV}, Appendix 11).  Therefore, to the extent that our solution is a simple oscillation of the density at a single wavelength less than $1/(2k_F)$ (see Appendix) the use of the local approximation $\mu(q) = (d\mu/dn)\delta n(q)$ is reasonable for $q<2k_F$, which is a much weaker condition than $q\ll 2k_F$.}

%
%

We conclude with a few comments on the order of magnitude of the predicted CDW and propose a possible experiment to confirm its existence.  The order of magnitude of the density modulation is
$\frac{\varepsilon F}{e^2}\simeq10^5$ cm$^{-2}$ for an electric field of 1 V/m (we assume a dielectric constant $\epsilon\sim 2$ for the environment and an effective electron mass approximately equal to the bare mass).  Thus for an equilibrium density  of order $10^{12}$ cm$^{-2}$ an electric field of order $10^5$ V/m produces a density modulation $1\%$ of the equilibrium density.  As for the wavelength of the CDW, a typical value of negative $\lambda$, deduced from Fig. 1, is $\lambda \simeq -0.05 (na)^{-1}\simeq  -0.05\mu$m.  

To realize the Coulomb drag experiment depicted in Fig. 2 we consider a trilayer structure formed by monolayers of MoSe$_2$, WSe$_2$, and graphene, with a band alignment shown in Fig. 5, which is predicted by first-principle calculations \cite{Guo2016} and experimentally confirmed~\cite{Yu2009}. Chemical vapor deposition and mechanical exfoliation techniques to fabricate such van der Waals multilayers are well established \cite{Wang2016,Ceballos2017}.  Steady illumination by a laser beam is used to globally inject electron-hole pairs in MoSe$_2$. The holes transfer to graphene, as dictated by the band alignment, while the electrons remain in MoSe$_2$ due to the energy barrier provided by the middle WSe$_2$ layer. This allows the graphene and MoSe$_2$ layers to be p- and n-doped, respectively, with equal carrier densities, so that global charge neutrality is preserved.  For carrier lifetimes of the order of nanoseconds, a carrier density of 10$^{12}$ cm$^{-2}$ is easily achieved. 

\begin{figure}[ht]
\begin{center}
\includegraphics[scale=0.75]{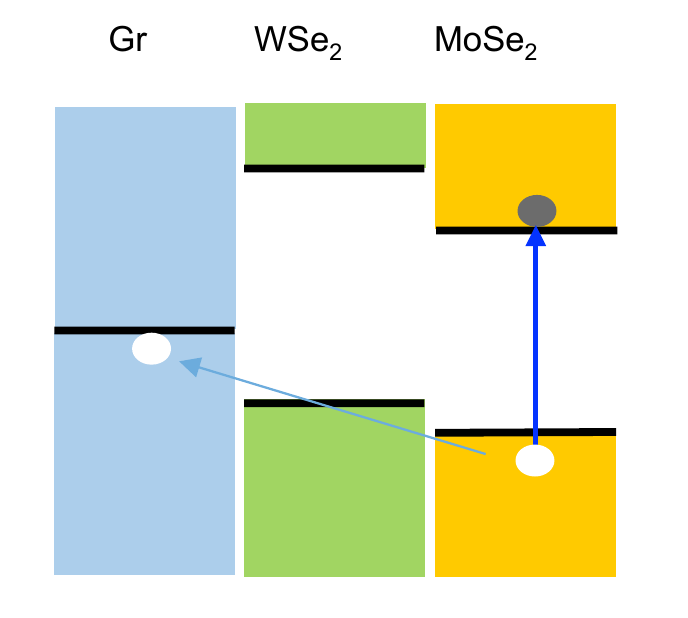}
\end{center}
\caption{Trilayer structure for the observation of electrically induced CDW.  Upon global illumination a hole gas is generated in graphene (Gr) and an electron gas in MoSe$_2$.  A hole current is driven in the graphene layer generating, via Coulomb drag a CDW in the MoSe$_2$ layer.}
\end{figure}

The large gap exhibited by the MoSe$_2$ layer not only allows to reach large doping concentrations without introducing disorder, but is also beneficial for the realization of the negative compressibility. Indeed, it is well known~\cite{Kotov2012} that in 2D semimetals, such as graphene, the compressibility  does not becomes negative because of the positive contribution of the completely-filled valence band. In contrast to this, in MoSe$_2$ the large gap allows the contribution of the conduction band to dominate and to change the sign of the compressibility for sufficiently low carrier concentration.

By applying a voltage of 10 V over a 10 $\mu$m graphene  channel, with a room temperature mobility of 10$^4$ cm$^2$/(V.s) \cite{Geim2013} a hole current density of 10$^{3}$ A/m  is generated in a $\mu$m-wide layer.   The target Coulomb drag potential of $10^5$ V/m in the MoSe$_2$ electron layer requires a Coulomb drag transresistivity of 10$^2$ ohm.  Such a value of the transresistivity seem to be realizable in van der Waals heterostructures~\cite{Geim2013}, in which the distance between the layers is only a few Angstroms.  Indeed, values of the transresistivity of the order of 50 ohm have been observed in experiments on bilayer graphene at temperatures of 240 K.\cite{Gorbachev2012}  The system we propose is expected to support larger transresistivity due to the larger effective mass and consequently higher density of states of electrons in MoSe$_2$. 
To evaluate the feasibility of optical detection of such a CDW, we note that in MoSe$_2$ a carrier density on the order of 10$^{12}$ cm$^{-2}$ changes the absorption coefficient by 10$^{-3}$ at the excitonic resonance.~\cite{Kumar2014} With proper modulation techniques, it has been demonstrated that a differential absorption (relative change of the absorption coefficient) of the order of 10$^{-7}$ can be detected.~\cite{Cui2015}  A 1\% CDW amplitude would yield a differential absorption signal of the order of 10$^{-5}$, which is two orders of magnitude higher than the detection sensitivity. For a CDW with a wavelength of the order of 1 $\mu$m, a direct imaging of the differential absorption can resolve the carrier density modulation. For shorter wavelength CDW, such as 100 nm, a spatial derivative technique can be used~\cite{Werake2011}. Alternatively, diffraction of a light from the formed grating can be used to detect the CDW.

The innovative experimental setup we propose offers several advantages with respect to standard ones based on semiconductor quantum wells and graphene heterostructures. First of all, the ``doping by illumination'' method has the advantage of keeping the system clean and free of metallic gates and other external sources of screening, thereby enhancing both the drag transresistivity and the negative contribution to the compressibility.
Moreover, the detection method allows to follow the evolution of the CDW wavelength with the density of carriers, allowing to distinguish the non-equilibrium CDW from other phenomena such as Friedel oscillations.

{\it Acknowledgments}-- We acknowledge support from NSF Grants DMR-1406568 (GV) and DMR-1505852 (HZ). One of us (GV) is indebted to Boris Shklovskii for many insightful comments and suggestions.

\begin{appendix}
\section{Appendix: Analytic solution of Eq.(\ref{IntegroDifferential}) }
After expressing the density modulation as 
\be
\delta n(x)=\frac{x}{\pi\sqrt{1-x^2}}+f(x)\,,
\ee
the equation (\ref{IntegroDifferential}) for $f(x)$ takes the form
\be
\int_{-1}^{+1}\frac{f(x')}{x-x'} dx' -\bar \lambda f'(x)=
\frac{\bar \lambda} {\pi (1-x^2)^{3/2}}\,,
\ee
where the integral is understood in the principal value sense.
We seek a solution in the form
\be
f(x)=f(q)\sin(qx)\,,
\ee
where $q>0$.
Making use of the identity
\be
\int_{-\infty}^{+\infty}\frac{\sin(qx')}{x-x'} dx'=-\pi \cos(qx)\,,
\ee
we rewrite the first term on the left hand side of Eq.(\ref{IntegroDifferential}) as
\be
\int_{-1}^{+1}\frac{\sin(qx')}{x-x'} dx'
=-\pi \cos(qx)-2\int_{1}^{\infty}\frac{x'\sin(qx')}{x^2-x'^2}dx'
\ee
and our equation for $f(q)$ becomes
\ber
&&\left\{(\pi+\lambda q)\cos(qx)+2\int_{1}^{\infty}\frac{x'\sin(qx')}{x^2-x'^2}dx'\right\}f(q)\nn\\
&=&-\frac{\bar \lambda} {\pi (1-x^2)^{3/2}}\,.
\eer
In the region $x\ll 1$ this simplifies to
\be
\left\{(\pi+\lambda q)\cos(qx)-2\int_{1}^{\infty}\frac{\sin(qx')}{x'}dx'\right\}f(q)=-\frac{\bar\lambda}{\pi}\,,
\ee
which is expressed in terms of the sine integral function, ${\rm Si}(q)\equiv\int_0^q \frac{\sin t}{t}dt$, as follows:
\be
\left\{(\pi+\lambda q)\cos(qx)-2\left(\frac{\pi}{2}-{\rm Si}(q)\right)\right\}f(q)=-\frac{\bar\lambda} {\pi}\,.
\ee
This equation requires
\be
\pi+\bar\lambda q=0~\rightarrow q=-\frac{\pi}{\bar\lambda}\,.
\ee
For $q>0$ a solution exists only if $\lambda$ is negative and 
\be
q=\frac{\pi}{|\bar\lambda|}\,.
\ee
Thus, the correction to the classical solution is 
\be
f(x)=-\frac{|\bar\lambda| \sin\left(\frac{\pi x}{|\bar\lambda|}\right)}{2\pi \left(\frac{\pi}{2}-{\rm Si}[\pi/|\bar\lambda|]\right)}\,.
\ee
In the limit $|\bar\lambda|\to0$ we have
\be
\frac{\pi}{2}-{\rm Si}[\pi/|\bar\lambda|]\simeq \frac{|\bar\lambda|}{\pi}\cos(\pi/|\bar\lambda|)\,,
\ee
yielding
\be
f(x)= - \frac{\sin\left(\frac{\pi x}{|\bar\lambda|}\right)}{2\cos(\pi/|\bar\lambda|)}\,.
\ee
Observe the strong non-analyticity of the solution for $|\bar\lambda|\to 0$.  While the wavelength of the density oscillations shrinks to zero, the derivative grows so that the amplitude remains constant in the limit.  As discussed in the main text, the limit $\bar \lambda \to 0$ is unphysical if interpreted as $\lambda \to 0$ at finite system size $L$, but has a clear physical significance for finite $\lambda$ and $L\to \infty$, where it describes finite amplitude oscillations in a system of infinite size.  

\end{appendix}


\end{document}